\documentclass[twocolumn,showpacs,amsmath,amssymb,superscriptaddress,aps]{revtex4-1}
\usepackage[dvips]{graphicx}
\usepackage{epsfig}
\usepackage{amsmath}
\usepackage[dvips]{color}
\newcommand{\beq}{\begin{eqnarray}}
\newcommand{\eeq}{\end{eqnarray}}
\newcommand{\oli}{\mbox{$^{11}{\rm Li}$}}
\newcommand{\lid}{\mbox{$^{11}{\rm Li}+d$}}
\newcommand{\lip}{\mbox{$^{11}{\rm Li}+p$}}
\newcommand{\lipb}{\mbox{$^{9}{\rm Li}+p$}}
\newcommand{\lin}{\mbox{$^{11}{\rm Li}+n$}}
\newcommand{\linb}{\mbox{$^{9}{\rm Li}+n$}}
\newcommand{\linn}{\mbox{$^{9}{\rm Li}+n+n$}}
\newcommand{\emax}{\mbox{$E_{\rm max}$}}

\newcommand{\jmax}{\mbox{$j_{\rm max}$}}
\newcommand{\kmax}{\mbox{$K_{\rm max}$}}
\newcommand{\elab}{\mbox{$E_{\rm lab}$}}
\newcommand{\ecm}{\mbox{$E_{\rm c.m.}$}}

\begin{document}
	\title{Low-energy $\lip$ and $\lid$ scattering in a multicluster model}
	\author{P. Descouvemont}
	\email{pdesc@ulb.ac.be}
\affiliation{Physique Nucl\'eaire Th\'eorique et Physique Math\'ematique, C.P. 229,
	Universit\'e Libre de Bruxelles (ULB), B 1050 Brussels, Belgium}
		\date{\today}
\begin{abstract}
The $\lip$ and $\lid$ reactions are investigated in the  Continuum Discretized Coupled Channel (CDCC) method
with a three-body description ($\linn$) of $\oli$. I first discuss the properties of $\oli$, and focus on E1 transition probabilities to
the continuum. The existence of a $1^-$ resonance at low excitation energies is confirmed, but the associated E1 transition from 
the ground state does not have an isoscalar character, as suggested in a recent experiment. In a second step, I study
the $\lip$ elastic cross section at $\elab=66$ MeV in the CDCC framework. I obtain a fair agreement with experiment, and show 
that breakup effects are maximal at large angles. The breakup cross section is shown to be dominated by the $1^-$ dipole state in
$\oli$, but the role of this resonance is minor in elastic scattering. 
From CDCC equivalent $\lip$ and $\lin$ potentials, I explore the $\lid$ cross section within a standard three-body
$\oli+(p+n)$ model. At small angles, the experimental cross section is close to the Rutherford scattering cross section,
which is not supported by the CDCC. A five-body ($^9{\rm Li}+n+n)+(p+n)$ calculation is then performed.
Including breakup states in $\oli$ and in the deuteron represents a numerical challenge for theory, owing to the large number
of channels. Although a full convergence could not be reached, the CDCC model tends to overestimate the data at 
small angles. I suggest that measurements of the $\lipb$ elastic scattering would be helpful to determine more accurate
optical potentials. The current disagreement between experiment and
theory on $\lid$ scattering also deserves new experiments at other energies.
\end{abstract}
\maketitle

\section{Introduction}
Since the discovery of halo nuclei \cite{THH85}, the physics of exotic nuclei attracted much interest in the nuclear 
physics community \cite{TSK13,SLY03}.  Halo nuclei are characterized by a low separation energy of the last nucleon(s), 
and therefore by an unusually large radius.  More generally, exotic nuclei, which are at the limit of stability, present 
short lifetimes, and are essentially studied through reactions.

The recent advances of radioactive beams opened many new perspectives in the physics of exotic nuclei.  In parallel 
it becomes more and more necessary to develop theoretical models, which can help in the interpretation of the data.  
Among the various theories, the Continuum Discretized Coupled Channel (CDCC) method \cite{Ra74,AIK87} is well suited for 
reactions involving exotic nuclei.  In the CDCC method, the continuum of the projectile is taken into account.  
This effect was first shown in $d+$ nucleus data, owing to the low binding energy of the deuteron \cite{Ra74}.  
The CDCC formalism was then successfully applied to reactions involving weakly bound nuclei such as $^{11}$Be \cite{DSM12} 
or $^6$He \cite{MHO04}.

The first variant of the CDCC method considered a two-body projectile (typically $d=p+n$) on a structureless target.  
More recently, it was extended to three-body projectiles \cite{MHO04}, and even to two-body projectile and target \cite{De18}.  
The present works addresses the $\lip$ and $\lid$ reactions, which have been experimentally investigated recently \cite{TKA17,KST15}.  
One of the main conclusions drawn in Refs.\ \cite{TKA17,KST15} is the presence of a dipole resonance in $\oli$.  The authors 
measured the elastic cross section, together with an inelastic cross section to a broad $1^-$ resonance of $\oli$.

The main goal of the present work is to analyze both systems in a common framework, i.e. with the same $\oli$ wave functions, and 
with the same $^9$Li+nucleon interaction.  For this purpose, I use a three-body $\linn$ model to describe $\oli$, and apply the 
CDCC theory for the scattering cross sections.  Using a common approach for $\lip$ and $\lid$, however, requires a generalization 
of the CDCC method to three-body + two-body systems.  This extension raises significant numerical difficulties, 
but can be performed with modern computing facilities.
The $\lip$ reaction was recently investigated by Matsumoto {\sl et al}.\ \cite{MTO19} in the CDCC formalism.  The authors, however, 
do not consider $\lid$ scattering and essentially focus on a possible Feshbach resonance in the $\linn$ system.

The text is organized as follows.  In Sec. II, I discuss the $\oli$ three-body model and, in particular, E1 transitions
to the continuum.  
Section III is devoted to the CDCC formalism which is presented in a way which is valid for any number of constituents.  
In Sections IV and V, I show the $\lip$ and $\lid$ cross sections, respectively.  I also discuss equivalent
potentials. The conclusion and outlook are presented in Sec. VI.

\section{Three-body description of $\oli$}
\subsection{Outline of the hyperspherical method}
The hyperspherical method is well adapted to three-body systems (see for example, Ref.\ \cite{ZDF93}), even for scattering 
states \cite{DTB06}.  I consider a three-body nucleus, made of a core (the spin is neglected) and of two neutrons.  
The nucleon number and the charge of the core are denoted as $A_1$ and $Z_1 e$, respectively.  There are three possible sets 
of scaled 
Jacobi coordinates ($\pmb{x},\pmb{y}  $) (see Refs.\ \cite{RR70,ZDF93} for more information).  I choose
\begin{eqnarray}
\pmb{x}=\frac{1}{\sqrt{2}}\left(\pmb{r}_3-\pmb{r}_2\right), 
\pmb{y}=\sqrt{\frac{2A_1}{A_1+2}}\left(\pmb{r}_1-\frac{\pmb{r}_2+\pmb{r}_3}{2}\right),
\label{eq1}
\end{eqnarray}
where $\pmb{r}_1,\pmb{r}_2 $ and $\pmb{r}_3$ are the coordinates of the core and of the neutrons.  This choice permits a 
natural symmetry of the wave functions regarding the exchange of the two neutrons.

In these coordinates, the $\oli$ Hamiltonian is given by
\begin{eqnarray}
H_{0}=-\frac{\hbar^2}{2m_N}\left(\Delta_x+\Delta_y\right)+\sum_{i<j}V_{ij},
\label{eq2}
\end{eqnarray}
where $m_N$ is the nucleon mass, and $V_{ij}$ are two-body potentials ($n+n$ and $\linb$).  The hyperadius $\rho$ and the hyperangle 
$\alpha$ are defined as
\begin{eqnarray}
\rho^2=x^2+y^2,\qquad
\alpha=\arctan\frac{y}{x},
\label{eq3}
\end{eqnarray}
and the wave function in angular momentum $j$ and parity $\pi$ is expanded as
\begin{eqnarray}
\Psi^{jm\pi}=\rho^{-5/2}\sum_{K=0}^{\infty}\sum_{\gamma} \chi^{j\pi}_{\gamma K}(\rho) {\cal Y}^{jm}_{\gamma K}(\Omega_{5\rho}).
\label{eq4}
\end{eqnarray}
In this definition, $K$ is the hypermoment, and $\gamma=(\ell_x,\ell_y,\ell,S)$ represents a set of quantum numbers \cite{DDB03}. 
The summation over $K$ is truncated at a maximum value $\kmax$. 
The hyperspherical harmonics ${\cal Y}^{jm}_{\gamma K}$ depend on five angles $\Omega_{5\rho}=(\Omega_x,\Omega_y,\alpha)$; 
they are defined in Ref.\ \cite{DDB03}.  In Eq.\ (\ref{eq4}), the hyperradial functions $\chi^{j\pi}_{\gamma K}(\rho)$ are obtained from 
a set of coupled differential equations
\begin{eqnarray}
&&\biggl( -\frac{\hbar^2}{2m_N}
\biggl[\dfrac{d^2}{d\rho^2}-\dfrac{{\cal L}_K({\cal L}_K+1)}{\rho^2}\biggr]-E \biggr)
\chi^{j\pi}_{\gamma K}(\rho) \nonumber\\
&&\hspace{1cm} +\sum_{K' \gamma'}V^{j\pi}_{\gamma K ,\gamma' K'}(\rho)\, \chi^{j\pi}_{\gamma' K'}(\rho)=0,
\label{eq5}
\end{eqnarray}
where ${\cal L}_K=K+3/2$ and where $V^{j\pi}_{\gamma K ,\gamma' K'}(\rho)$ are the coupling potentials, determined from the 
matrix elements of the two-body potentials in (\ref{eq2}) between hyperspherical harmonics.  

In the present work, I am looking for square-integrable solutions of Eq.\ (\ref{eq5}).  I expand the hyperradial functions over a set of $N$ basis functions $u_i(\rho)$ as 
\begin{eqnarray}
\chi^{j\pi}_{\gamma K}(\rho)=\sum_{i=1}^N c^{j\pi}_{\gamma  K i}u_i(\rho).
\label{eq6}
\end{eqnarray}
In practice, I choose Lagrange functions \cite{Ba15} which allow a simple and accurate calculation of matrix elements, and 
which have been used in previous works \cite{PDB12,DDC15}.

\subsection{Description and properties of $\oli$}

The $n+n$ potential is the central part 
of the Minnesota potential with the exchange parameter $u=1$ \cite{TLT77}. The $\linb$ potential is chosen as in Ref.\ \cite{EBH97}, and reproduces 
various properties of $^{10}$Li, such as the scattering length.  
Notice that this potential contains a forbidden state for the $s$ and $p_{3/2}$ partial waves.  To avoid spurious three-body 
states in the solution of (\ref{eq5}), a supersymmetric transformation \cite{Ba87} is applied.  As in Ref.\ \cite{PDB12}, I scale the $\linb$ potential by a factor $1.0051$ 
to reproduce the $\oli$ two-neutron binding energy $S_{2n}=0.378$ MeV \cite{BAG08}. 
For the basis functions $u_i(\rho)$, I use a Gauss-Laguerre mesh with $N=20$ and a scale parameter $h=0.3$ fm. I adopt
$\kmax=20$ for the ground state, which guarantees the convergence of the energy and of the r.m.s. radius.
Using a $^9$Li radius of 2.43 fm \cite{EAA02}, 
I find a $\oli$ r.m.s radius of 3.12 fm, in fair agreement with experiment  $3.16 \pm 0.11$ fm \cite{TKY88}.

The structure and the E1 distribution of $\oli$ have been previously discussed \cite{PDB12}.  In the present work, I want 
to address the E1 distribution more precisely.  A recent experimental work on $\lip$ scattering \cite{TKA17} suggests the existence 
of a low-energy $1^-$ resonance in $\oli$, and that the E1 transition probability to the ground state should have an isoscalar character.  According to the authors, this property arises from the halo structure of $\oli$.

Let me start with a microscopic interpretation of the E1 transitions.  At the long-wavelength approximation, the E1 operator is given,
in a $A$-nucleon model, by
\begin{eqnarray}
{\cal M}^{E1}_{\mu}=e\sum_{i=1}^A \bigl(\frac{1}{2}-t_{iz}\bigr)r'_i Y_1^{\mu}(\Omega'_i ),
\label{eq7}
\end{eqnarray}
where $\pmb{r}'_i=\pmb{r}_i-\pmb{R}_{c.m.}$, $\pmb{r}_i$ being the space coordinate of nucleon $i$, and $\pmb{R}_{c.m.}$ the 
center-of-mass coordinate.  Subtracting the c.m. coordinate ensures the Galilean invariance of the operator.  In this microscopic description, 
$t_{iz}$ is the isospin projection ($t_{iz}=+1/2$ for neutrons and $t_{iz}=-1/2$ for protons).  The first term is called 
isoscalar (it does not depend on isospin) and exactly vanishes for E1 transitions.  The second term is the isovector 
contribution which is essentially responsible for E1 transitions.  If the isospin of the initial and final states is $T=0$, 
the isovector term also vanishes.  A typical example is $^{16}$O and the $^{12}{\rm C}(\alpha,\gamma)^{16}$O  reaction.  
Using the long wavelength approximation with $T=0$ wave functions provides exactly zero for E1 transitions.  These transitions, 
however, play an important role in the capture reaction, and are due to small $T=1$ components \cite{DB87c}.

When I adapt definition (\ref{eq7}) to a non-microscopic model involving $N_C$ clusters with charges $Z_k$, I have
\begin{eqnarray}
{\cal M}^{E1}_{\mu}=e\sum_{k=1}^{N_C} Z_k r'_k Y_1^{\mu}(\Omega'_k ),
\label{eq8}
\end{eqnarray}
where $\pmb{r}'_k$ are now defined from the space coordinates of the clusters.  In a two-cluster model with 
nucleon numbers ($A_1,A_2$)
and charges ($Z_1e,Z_2e$), this leads to 
the well-known definition
\begin{eqnarray}
{\cal M}^{E1}_{\mu}=e \biggl( \frac{Z_1}{A_1}-\frac{Z_2}{A_2}\biggr)
r Y_1^{\mu}(\Omega_{r}),
\label{eq9}
\end{eqnarray}
where $\pmb{r}$ is the relative coordinate between the clusters.  In the present three-body model, involving a core and two neutrons, 
the dipole operator is, at the long wavelength approximation \cite{DDB03},
\begin{eqnarray}
{\cal M}^{E1}_{\mu}=e Z_1 \sqrt{ \frac{2}{A_1(A_1+2)}}
y Y_1^{\mu}(\Omega_y).
\label{eq10}
\end{eqnarray}
This form does not explicitly include isoscalar and isovector terms, as this wording is specific to the microscopic definition 
(\ref{eq7}).  However, the operator (\ref{eq10}) is directly deduced from (\ref{eq7}) and, therefore, is associated with an isovector contribution,
since the isoscalar term vanishes.

The E1 transition probabilities between an initial state $J_i\pi_i$ and a final state $J_j\pi_f$ are defined as
\begin{eqnarray}
&&B(E1,J_i\pi_i \rightarrow J_f\pi_f n)=\nonumber \\
&&\hspace{1 cm}\frac{2J_f+1}{2J_i+1} \vert 
\langle \Psi^{J_j\pi_f n}\Vert {\cal M}^{E1} \Vert \Psi^{J_i\pi_i}\rangle \vert^2.
\label{eq10b}
\end{eqnarray}
As in Ref.\ \cite{PDB12}, 
a smooth E1 distribution $dB(E1)/dE$ is obtained by folding the discrete $B(E1)$ with a Gaussian function centered at $\ecm$
(the width is $\sigma=0.3$ MeV, close to the 
experimental resolution). Here $J_i\pi_i$ corresponds to the ground state, and $J_j\pi_f n$ to the final pseudostates.
The E1 distribution computed with (\ref{eq10}) has been presented in Ref.\ \cite{PDB12}.  It presents a 
peak around $\ecm=0.5$ MeV, associated with a dipole resonance, and is qualitatively in agreement with the experimental data \cite{NVS06}
($\ecm$ is the energy with respect to the $\linn$ threshold).  

Recently, Tanaka {\sl et al}.\ \cite{TKA17} have measured the elastic and inelastic $\lip$ cross sections at $\elab=66$ MeV.  This work 
suggests a dipole resonance at $E_x=0.8$ MeV, in good agreement with the theoretical prediction.  According to Tanaka {\sl et al.}, 
the dipole transition should have an isoscalar component, due to the halo nature of $\oli$.  I have tested this hypothesis 
by extending the definition of the E1 operator beyond the long wavelength approximation.  The E1 operator then reads, in a microscopic 
model \cite{LC82},
\begin{eqnarray}
{\cal M}^{E1}_{\mu}&=&e\sum_{i=1}^A \bigl(\frac{1}{2}-t_{iz}\bigr)r'_i Y_1^{\mu}(\Omega'_i )
\biggl(1-\frac{(k_{\gamma}r'_i)^2}{10}\biggr) \nonumber \\
&&+i \frac{ek_{\gamma}}{4m_N c} \sum_{i=1}^A \bigl(\frac{1}{2}-t_{iz}\bigr)r'_i Y_1^{\mu}(\Omega'_i )
\pmb{r}'_i \cdot \pmb{p}'_i,
\label{eq11}
\end{eqnarray}
where $\pmb{p}_i$ is the momentum of nucleon $i$, and $k_{\gamma}$ is the photon momentum (a spin-dependent term is
neglected).
With this generalization, isoscalar transitions are possible.  However they are not expected to be important since $(k_{\gamma}r_i)^2$ is small (the photon energies are of the order of a few MeV).  
I have estimated the isoscalar component in the present three-body model, 
with the first term of definition (\ref{eq11}).  
In hyperspherical coordinates, the dipole operator (\ref{eq10}) is therefore complemented by an additional term
\begin{eqnarray}
{\cal M}^{E1,{\rm add.}}_{\mu}=-\frac{1}{10}
\biggl(\sqrt{ \frac{2}{A_1(A_1+2)}} k_{\gamma} y\biggr)^2 {\cal M}^{E1}_{\mu}
\label{eq12}
\end{eqnarray}

\begin{figure}[htb]
	\begin{center}
		\epsfig{file=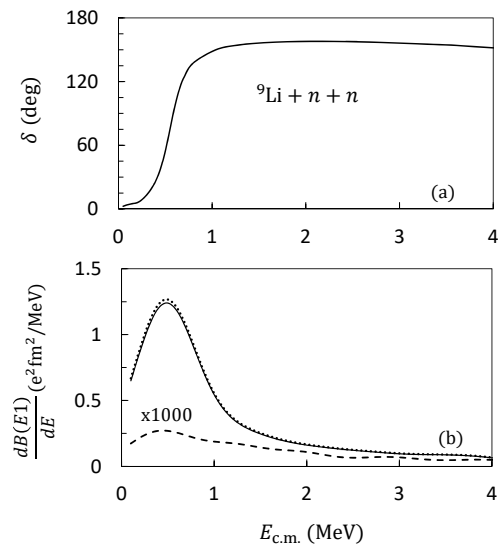,width=6.5cm}
		\caption{$\oli$ $1^-$ phase shift (a) (see Ref.\ \cite{PDB12}) and E1 distribution (b). In (b), the dashed 
			line represents the contribution of Eq.\ 
			(\ref{eq12}) only. The solid and dotted lines are obtained with the full E1 operator, and with the long-wavelength approximation
		(\ref{eq10}), respectively. $\ecm$ is the energy with respect to the $\linn$ threshold.}
		\label{fig_e1}
	\end{center}
\end{figure}

Figure \ref{fig_e1}(a) presents the $1^-$ $\linn$ three-body phase shift, as calculated in Ref.\ \cite{PDB12}.  
The E1 distribution is shown in Fig.\ \ref{fig_e1}(b) with the first-order term (dotted line) and with the full operator (solid line).  
The presence of a dipole resonance around $\ecm=0.5$ MeV is consistent in both figures, and seems well established from 
experiment (breakup \cite{NVS06} and inelastic scattering \cite{TKA17}).  However, Fig.\ \ref{fig_e1}(b) shows that the contribution 
of higher-order terms (dashed line) is negligible.  The present three-body calculation therefore confirms the existence of a dipole 
resonance at low energies (let me emphasize that the only parameter, a scaling factor of the $\linb$ potential, is adjusted on the 
ground-state energy), but does not support the interpretation of an isoscalar character.  This is not surprising since the next-order 
term is proportional to $(k_{\gamma}r)^2$.  Even if typical radii of halo nuclei are larger than in stable nuclei, the factor 
$k_{\gamma}=E_{\gamma}/\hbar c$ is of the order of 0.01 fm$^{-1}$, and makes the correction quite small.  Of course this 
argument could not be true in $T=0$ nuclei, since the leading term of the E1 operator exactly vanishes.

\section{Brief overview of the CDCC theory}
Originally, the CDCC method has been developed to describe $d+$nucleus scattering \cite{Ra74}.  Owing to the low breakup threshold 
of the deuteron, elastic scattering cannot be satisfactorily described if breakup effects are neglected.  
The basic idea of the CDCC 
method is to simulate breakup effects by approximations of the deuteron continuum, referred to as pseudostates (PS).  These PS correspond 
to positive eigenvalues of the Schr\"odinger equation associated with the projectile.  They do not have a specific physical meaning 
but represent an approximation of the continuum.  The CDCC formalism was very successful to reproduce various $d+$nucleus data.

With the advent of radioactive beams, the CDCC method turned out to be a useful tool to analyse reactions involving exotic 
nuclei \cite{YOM12}.  As for the deuteron, the neutron or proton separation energy of exotic nuclei is low, 
and breakup effects are expected to play an important role in reactions.
The original CDCC formalism was developed for two-body projectiles on structureless targets \cite{KYI86,AIK87}.  
This is well adapted to the scattering 
of typical two-body nuclei, such as $d$, $^7$Li, $^{11}$Be on heavy targets.  The formalism was then extended to three-body 
projectiles such as $^6$He \cite{MHO04} or $^9$Be \cite{DDC15}, and to systems involving two-body projectile and target, such as 
$^{11}{\rm Be}+d$ \cite{De18}. 

The goal of the present work is to analyse recent $\lip$ and $\lid$ data \cite{TKA17,KST15} in the CDCC
framework.  A realistic description 
of $\oli$ requires a three-body $\linn$ model, as discussed in Sect. II.  On the other hand, the $\lid$ reaction also involves the 
deuteron, which should be described by a $p+n$ structure.  Previous calculations on $^{11}{\rm Be}+d$ in a four-body CDCC 
model \cite{De17,De18} have shown that these calculations lead to a large number of channels (up to several thousands), but provide 
an excellent description of elastic scattering.

\begin{figure}[htb]
	\begin{center}
		\epsfig{file=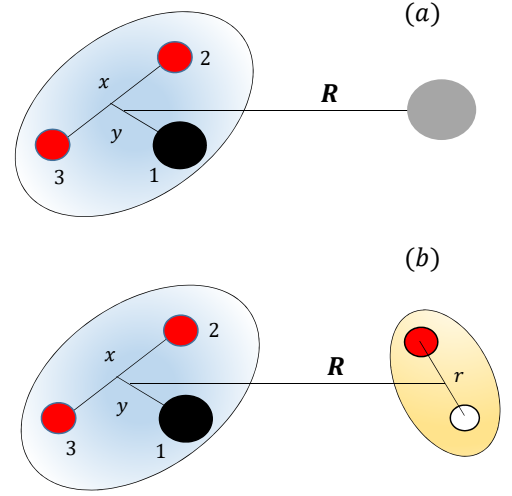,width=6.5cm}
		\caption{Cluster configurations and coordinates for $3+1$ (a) and $3+2$ (b) systems.}
		\label{fig_conf}
	\end{center}
\end{figure}

Let me consider a system of two nuclei described by a set of internal coordinates $\pmb{\xi}_i$ (see Fig.\ \ref{fig_conf}), and by an internal 
Hamiltonian $H_i$.
For a two-body system, I have 
\begin{eqnarray}
&&\pmb{\xi}_i=\pmb{r}_i, \nonumber \\
&&H_i=-\frac{\hbar^2}{2\mu_i}\Delta_i+v_{12}(r_i),
\end{eqnarray}
where $\mu_i$ is the reduced mass and $v_{12}(r_i)$ a (real) nucleus-nucleus potential.
In a three-body system
\begin{eqnarray}
\pmb{\xi}_i&=&(\pmb{x},\pmb{y}),\nonumber \\
H_{i}&=&-\frac{\hbar^2}{2m_N}\left(\bigtriangleup_x+\bigtriangleup_y\right)+v_{12}(x)+v_{13}(x,y)\nonumber \\
&&+v_{23}(x,y). 
\label{eq14}
\end{eqnarray}

The starting point of all CDCC calculations is to solve the Schr\"{o}dinger equation associated with the colliding nuclei, i.e.
\begin{eqnarray}
H_i\, \Phi^{jm\pi}_{k}=E^{j\pi}_{k} \, \Phi^{jm\pi}_{k},
\label{eq15}
\end{eqnarray}
where 
\begin{eqnarray}
\Phi^{jm\pi}_{k}&=&r^{-1}g^{j\pi}_{\ell k}(r)\, \bigl[ Y_{\ell}(\Omega)\otimes \chi^s\bigr]^{jm} \ {\rm for \ a \ 2-body \ system} \nonumber \\
&=&\rho^{-5/2}\sum_{\gamma K}\chi^{j\pi}_{\gamma K k}(\rho) {\cal Y}^{jm}_{\gamma K}(\Omega_{\rho}) \ {\rm for \ a \ 3-body \ system} \nonumber \\
\label{eq16}
\end{eqnarray}
In (\ref{eq15}), index $k$ refers  to the excitation level.  Energies with $E^{j\pi}_{k}<0$ correspond to physical states, 
and $E^{j\pi}_{k}>0$ correspond to PS.  

Let me now consider the Hamiltonian of the projectile + target system, which reads
\begin{eqnarray}
H=H_1+H_2+T_R+\sum_{ij}V_{ij}(\pmb{R},\pmb{\xi}_1,\pmb{\xi}_2),
\label{eq17}
\end{eqnarray}
where $\pmb{R}$ is the relative coordinate (see Fig. 2) and $V_{ij}$ are optical potentials between the fragments.  In $\lip$, I 
need $\lipb$ and $n+p$ optical potentials, whereas $\lid$ require the additional $^9{\rm Li}+n$ and $n+n$ potentials.  
The total wave function is expanded over a set of PS as
\begin{eqnarray}
\Psi^{JM\pi}=\sum_{c LI} u^{J\pi}_{c LI}(R)\,
\varphi^{JM\pi}_{c LI} (\Omega_R,\pmb{\xi}_1,\pmb{\xi}_2),
\label{eq18}
\end{eqnarray}
where index $c$ stands for $c=(j_1,k_1,j_2,k_2)$, $L$ is the relative angular momentum and $I$ the channel spin.
The channel functions $\varphi^{JM\pi}_{c LI}$ are defined from the internal wave functions of the 
projectile and target as
\begin{eqnarray}
&&\varphi^{JM\pi}_{c LI} (\Omega_R,\pmb{\xi}_1,\pmb{\xi}_2)= \nonumber \\
&& \hspace*{1cm}
\biggl[ \bigl[ 
\Phi^{j_1}_{k_1}(\pmb{\xi}_1)\otimes \Phi^{j_2}_{k_2}(\pmb{\xi}_2)\bigr]^I
\otimes Y_L(\Omega_R)\biggr]^{JM}.
\label{eq19}
\end{eqnarray}
The summations over the spins $j_1,j_2$ and over the excitation levels $k_1,k_2$ are controlled by truncation parameters
$\jmax$ and $\emax$ (which can be different for the target and for the projectile).
The radial functions $u^{J\pi}_{c LI}(R)$ in (\ref{eq18}) are obtained from a set of coupled equations
\begin{eqnarray}
&&\biggl[-\frac{\hbar^2}{2\mu}\biggl(\frac{d^2}{dR^2}-\frac{L(L+1)}{R^2}  \biggr)
+E^{j_1}_{k_1}+E^{j_2}_{k_2}-E \biggr]u^{J\pi}_{c LI}(R)\nonumber \\
&&\hspace{1cm}+\sum_{c'L'I'}V^{J\pi}_{cLI,c'L'I'}(R)\, u^{J\pi}_{c' L'I'}(R)=0,
\label{eq20}
\end{eqnarray}
where $\mu$ is the reduced mass, and where the coupling potentials $V^{J\pi}_{cLI,c'L'I'}(R)$ are defined from 
the matrix elements
\begin{eqnarray}
V^{J\pi}_{cLI,c'L'I'}(R)=\langle \varphi^{JM\pi}_{c LI} \vert \sum_{ij}V_{ij} \vert
\varphi^{JM\pi}_{c' L'I'}\rangle,
\label{eq21}
\end{eqnarray}
and involve integration over $\pmb{\xi}_1,\pmb{\xi}_2$ and $\Omega_R$.  The calculation of these coupling potentials is simple 
for two-body projectiles on structureless targets (see, for example, Ref.\ \cite{DBD10}).  For 3-body projectiles, 
the calculations are far more complicated \cite{MHO04}.  Here, I 
still go beyond this situation, since I consider a three-body projectile $\oli=\linn$ on a two-body target $d=p+n$.  
Some technical information is given in the Appendix.

The most challenging part of CDCC calculations, however, is not the calculation of the coupling potentials.  
The main problem is that system (\ref{eq20}) 
may involve several thousands of coupled equations, and must be solved for each $J\pi$.  In practice I use the $R$-matrix method 
with a Lagrange mesh \cite{DB10,De16a}.  This approach provides the scattering matrices, and therefore the scattering cross sections.

\section{The $\lip$ scattering}

\subsection{Conditions of the calculation}

The coupled-channel system (\ref{eq20}) is solved with the $R$-matrix method and Lagrange functions \cite{DB10,De16a}.  Typically 
I use a channel radius $a=25$ fm with 50 basis functions.  Small variations of these conditions do not bring 
any significant change in the cross sections.  I use the Koning-Delaroche potential \cite{KD03} (referred to as KD)
for $^9{\rm Li}+p$, and the Minnesota 
interaction \cite{TLT77} for $n+p$.  Of course the KD global potential is not fitted on $^9{\rm Li}+p$ data, which do no exist, 
and is therefore not expected to provide an excellent description of $\lip$.  To assess the sensitivity of the cross sections, 
I also use the Chapel Hill \cite{VTM91} parametrization (referred to as CH) for $^9{\rm Li}+p$.  The Coulomb potential is treated exactly.  
In contrast with Ref.\ \cite{MTO19}, who use the JLM potential, I do not introduce any renormalization factor in the 
coupling potentials.  

\begin{figure}[htb]
	\begin{center}
		\epsfig{file=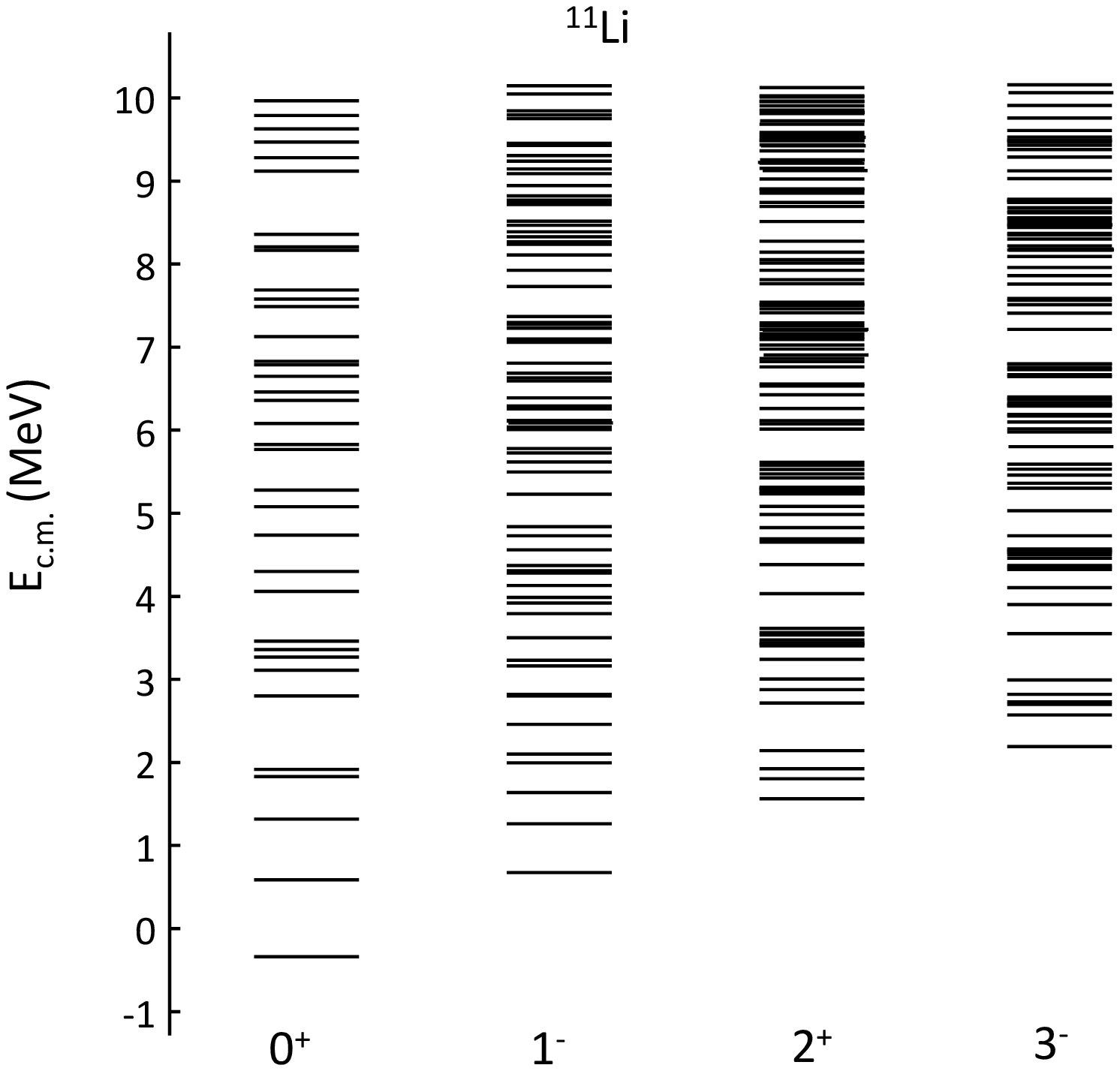,width=7.5cm}
		\caption{Pseudostate energies of $\oli$ for $j=0^+,1^-,2^+,3^-$ with the Lagrange basis defined in Sec. II.B.}
		\label{fig_spec}
	\end{center}
\end{figure}

The $\oli$ wave functions are described in Sec.\ II.  I include pseudostates for $j=0^+,1^-,2^+,3^-$ up to a maximum energy 
$\emax=10$ MeV, which provides an excellent convergence. The $\kmax$ values are $20,17,14,13$, respectively (the number
of components in (\ref{eq4}) rapidly increases with $j$ and $\kmax$). 
These states are illustrated in Fig.\ \ref{fig_spec}.  As in all CDCC calculations, only bound states (and narrow resonances) can 
be associated with physical states.  Other states are used to simulate the $\linn$ continuum, and depend on the choice of the basis.  
Converged calculations, however, should not depend on the $\oli$ basis. 

\subsection{$\lip$ cross sections}
\begin{figure}[htb]
	\begin{center}
		\epsfig{file=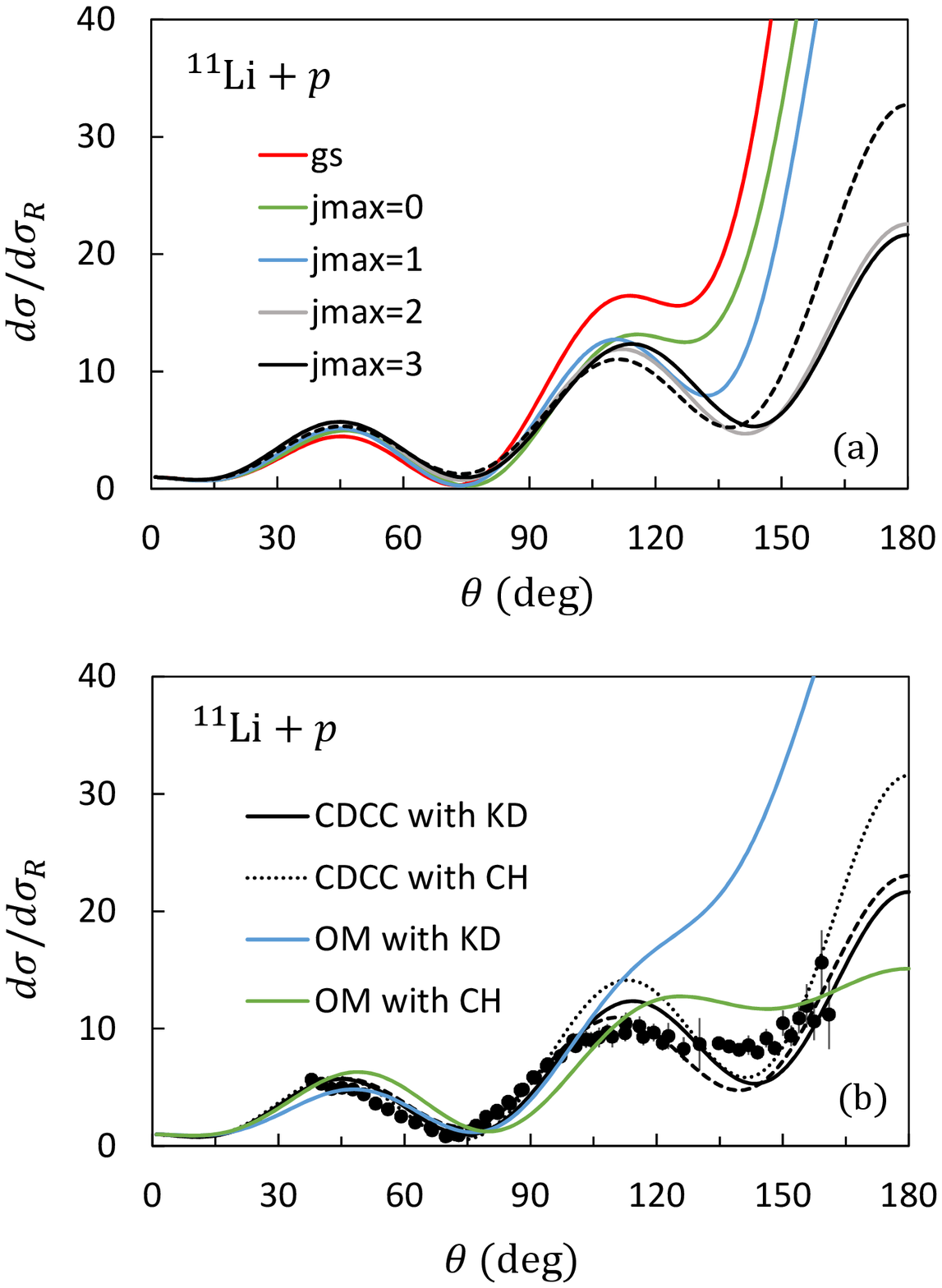,width=6.5cm}
		\caption{$\lip$ cross sections divided by the Rutherford cross sections at $\elab=66$ MeV ($\ecm=5.5$ MeV). (a) Convergence 
			with respect to $\jmax$. The dashed line corresponds to the truncation energy $\emax=5.5$ MeV, where closed channels are neglected.
			(b) Comparison of the CDCC cross sections with optical model (OM) calculations using the KD and CH $\lip$ potentials.
		The dashed line is obtained with a smaller $\oli$ basis (see text). The experimental data are taken from
	Ref.\ \cite{TKA17}.}
		\label{fig_sig_lip}
	\end{center}
\end{figure}

The convergence of the 
elastic cross section at $E_{\rm Li}=66$ MeV ($\ecm=5.5$ MeV) is illustrated in Fig.\ \ref{fig_sig_lip}(a). A linear scale
is used to highlight the differences between the calculations.  
For $\theta \lesssim 90^{\circ}$, the cross section is weakly sensitive to breakup effects.  At large angles, however, 
the single-channel calculation, involving the $\oli$ ground state only, provides large cross sections, in contradiction with experiment 
(see Fig.\ \ref{fig_sig_lip}(b)). Including more $\oli$ partial waves reduces the cross section at large angles.  
The $2^+$ pseudostates play the dominant role, whereas $j=1^-$ is less important.  
The cross sections for $\jmax=2$ and $\jmax=3$ are almost superimposed, which 
shows that the calculation is converged.  

The dashed line in Fig.\ \ref{fig_sig_lip}(a) is obtained with $\jmax=3$, but 
limiting the pseudostates to $\emax=5.5$ MeV.  In other words, open channels only are included in the expansion (\ref{eq18}).  
At large angles, the role of closed channels is therefore not negligible, as shown by the solid and dashed black curves 
(both with $\jmax=3$).  This confirms the conclusion of Ref.\ \cite{OY16}, i.e. that closed channels cannot be neglected 
to obtain converged cross sections.

I compare the CDCC results with experiment in Fig.\ \ref{fig_sig_lip}(b).  The solid black curve is the same as in Fig.\ \ref{fig_sig_lip}(a).  
With the same conditions ($\jmax=3,\emax=10$ MeV), I test the influence of two other inputs of the calculation.  
The dashed line is obtained with the $\lipb$ CH interaction \cite{VTM91}.  None of the available global parametrization 
provides $\lipb$ precise potentials and, strictly speaking, should not be used for light nuclei such as $^9$Li.  However, 
since no scattering data exist, the tradition in the literature is to use these compilations.  The comparison between 
KD and CH illustrates the precision that I may expect from the choice of the $\lipb$ interaction.  

The other input of the calculation is the $\oli$ basis.  In order to reduce the computer times, I have used a smaller basis for $\oli : N=15,h=0.25$ fm.  
This does not affect the $\oli$ ground state, but changes the continuum spectrum shown in Fig.\ \ref{fig_spec} (the density is lower).  
Keeping $\jmax=3$ and $\emax=10$ MeV provides the dashed line of Fig.\ \ref{fig_sig_lip}(b).  Again the effect is weak 
(a few percents at maximum) and shows up for $\theta > 100^{\circ}$ only. This smaller basis will be used for the
five-body $\lid$ calculations, where reducing the number of PS is a critical issue.

\begin{figure}[htb]
	\begin{center}
		\epsfig{file=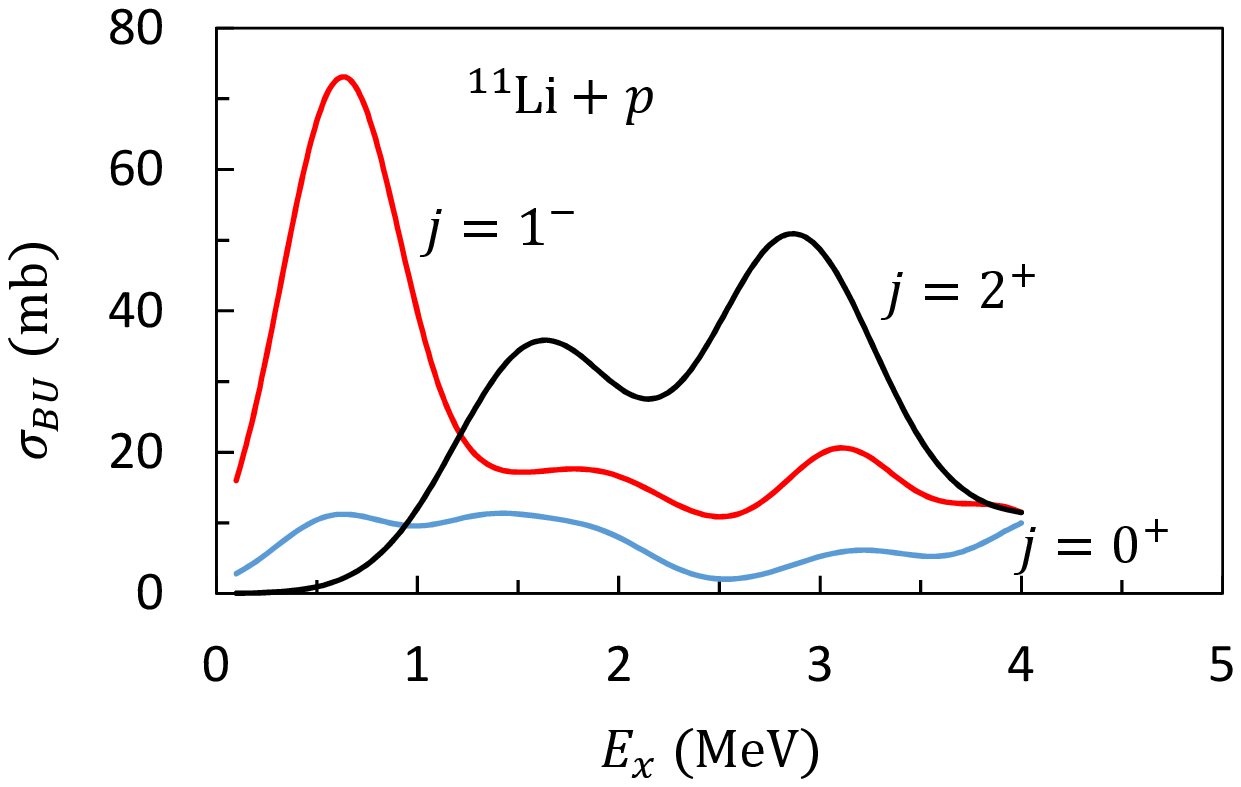,width=6.5cm}
		\caption{$\lip$ breakup cross sections at $\elab=66$ MeV ($\ecm=5.5$ MeV) for different $j$ values.}
		\label{fig_bu}
	\end{center}
\end{figure}

In Ref.\ \cite{TKA17}, the existence of a dipole resonance in $\oli$ was suggested from an inelastic measurement 
$\oli (p,p')$.  Excitation functions in different angular ranges show a peak around $E_x\approx 0.8$ MeV.  From a theoretical point 
of view, however, since the $1^-$ resonance discussed in Sec.\ II is quite broad, an inelastic cross section cannot be defined 
rigorously.  Consequently, in order to provide a link between the maximum in the $E1$ distribution (see Fig.\ \ref{fig_e1}) and the 
scattering process, I have computed the integrated breakup cross section.  The breakup cross section to a pseudostate $n$ is defined from
the scattering matrices as
\beq
\sigma_{\rm BU}^n(E,E_n)=\frac{\pi}{k^2}\sum_{J\pi}(2J+1)\sum_L \vert U^{J\pi}_{\omega,nL}\vert^2,
\label{eq22}
\eeq
where $\omega$ is the entrance channel, and $L$ is the relative angular momentum, which may take several values for pseudostates 
with $j>0$.  Equation (\ref{eq22}) gives the breakup cross section to a specific pseudostate at the breakup energy $E_n$.  To derive a 
smooth curve, I use a standard folding method \cite{PDB12}  with a Gaussian factor $f(E_x,E_n)$ (the width $\sigma$ is chosen as $\sigma=0.3$ MeV
 \cite{PDB12}).  This leads to the total cross section
\beq
\sigma_{\rm BU}(E,E_x)=\sum_n f(E_x,E_n) \sigma_{\rm BU}^n(E,E_n),
\label{eq23}
\eeq
where $E_x$ is the $\linn$ three-body energy.  The breakup cross section (\ref{eq23}) is shown in Fig.\ \ref{fig_bu}, where I separate 
the contributions of the different partial waves in $\oli$.  As expected from the $E1$ distribution of Fig.\ \ref{fig_e1}, 
the breakup cross section presents a maximum near $E_x=0.8$ MeV for $j=1^-$.  This is supported by the experimental inelastic 
cross sections of Ref.\ \cite{TKA17}.  The $j=0^+$ contribution is small but the $j=2^+$ component is dominant for $E_x\gtrsim 2$ MeV. The $2^+$ three-body phase shift presents a broad structure between 1 and 4 MeV \cite{PDB12}. In the discretized continuum approximation, this structure shows up as broad peaks, which are visible in Fig.\ \ref{fig_bu} but which do not correspond to
physical states.

\subsection{$\lip$ and $\lin$ equivalent potentials}

As mentioned before, CDCC calculations involve a large number of channels.  It is, however, possible to simulate these large-scale 
calculations by equivalent optical potentials.  The procedure follows Refs.\ \cite{TNL89,De18}, and provides potentials which 
approximately reproduce the multichannel CDCC calculations.  Having this equivalent optical potential, it is important to 
assess its accuracy to reproduce the CDCC elastic cross section.

\begin{figure}[htb]
	\begin{center}
		\epsfig{file=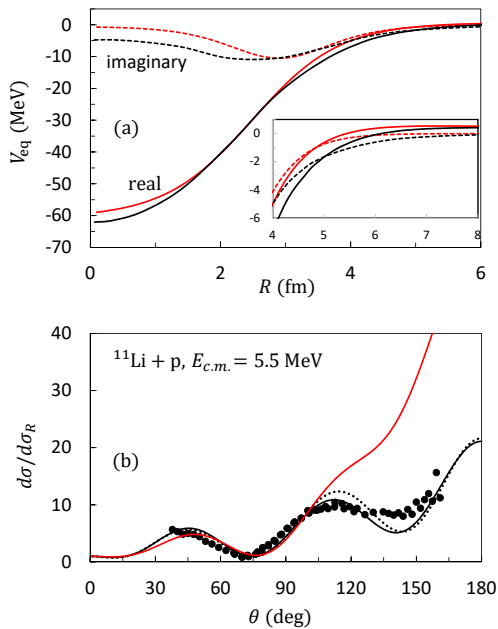,width=6.5cm}
		\caption{(a) Real and imaginary equivalent $\lip$ potentials with the CDCC (black curves). The KD potentials are
		shown in red. (b) Corresponding cross sections obtained with the potentials (solid lines) and with the full CDCC
	calculations (dotted line). The experimental data are taken from Ref.\ \cite{TKA17}.}
		\label{fig_pot_lip}
	\end{center}
\end{figure}

The $\lip$ equivalent potential $V_{\rm eq}(R)$ is shown in Fig.\ \ref{fig_pot_lip}(a), where I also plot the KD potential for the sake of 
comparison.  The general shapes of the real and imaginary terms are similar for both potentials.  The real component is a 
typical volume term, and the imaginary part corresponds to a surface absorption.  The inset of Fig.\ \ref{fig_pot_lip}(a) 
focuses on radial distances near the barrier, where the sensitivity of the cross section is the largest.  As expected, 
the role of breakup channels in CDCC is to reduce the barrier, and to increase the absorption at the surface.  This effect 
can be seen in Fig.\ \ref{fig_pot_lip}(b), where the cross sections are presented.  Although this global
parametrization is not fitted on exotic nuclei such as $\oli$, the calculation with the $\lip$ KD global potential 
reproduces fairly well the data up to $\theta\approx 100^{\circ}$, but strongly deviates at large angles.  
The cross section obtained with the equivalent CDCC potential 
(solid curve) is in excellent agreement with the full CDCC calculation (dotted curve), which shows that the potential of 
Fig.\ \ref{fig_pot_lip}(a) provides a good approximation of the CDCC model.

I complement this study with the $\lin$ scattering at the same energy.  The goal is twofold: (i) to test the KD global 
potential for neutrons; (ii) to define a $\lin$ equivalent potential which, together with the $\lip$ potential discussed before, 
will be used to investigate $\lid$ scattering.

\begin{figure}[htb]
	\begin{center}
		\epsfig{file=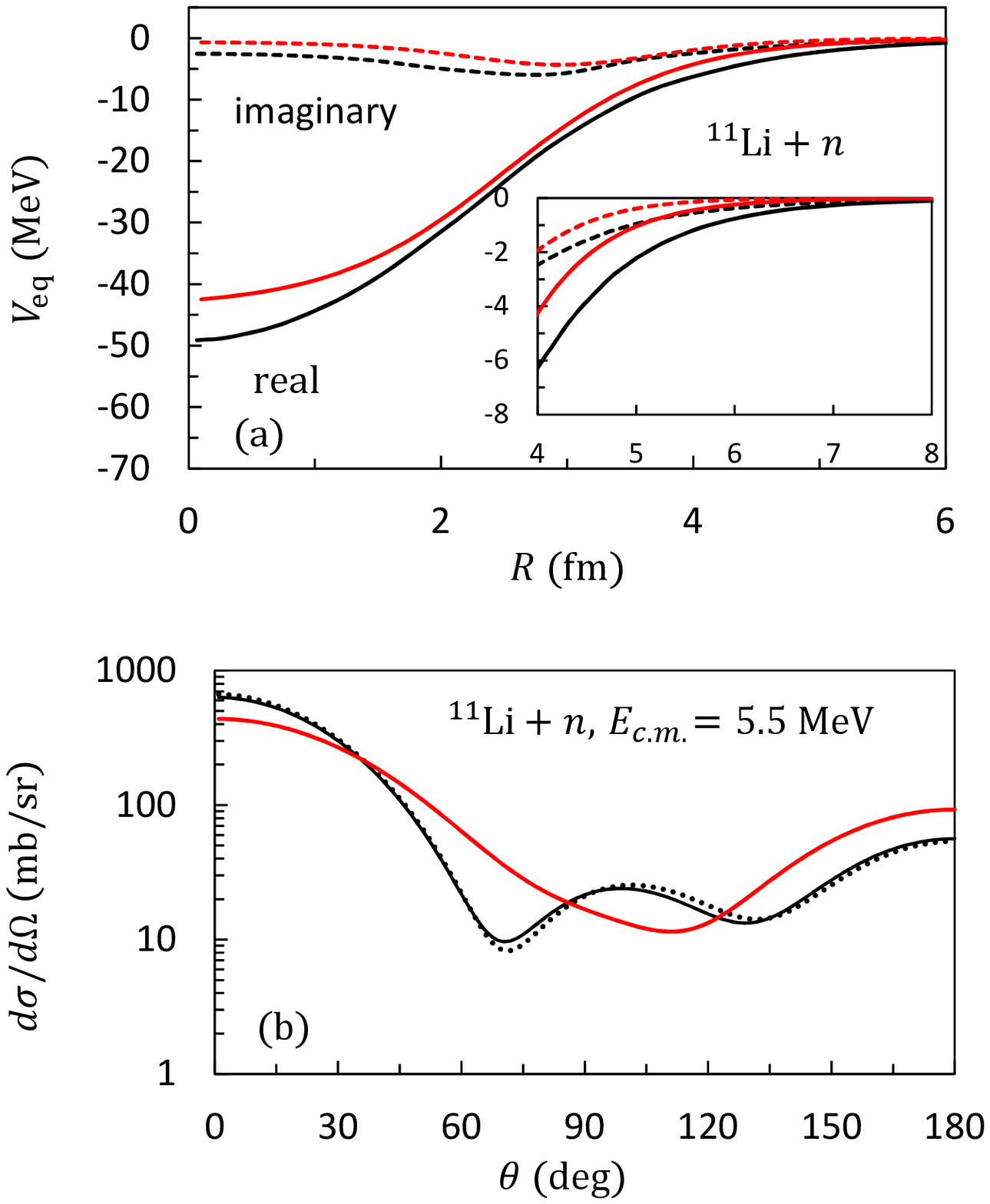,width=6.5cm}
		\caption{See caption to Fig.\ \ref{fig_pot_lip} for $\lin$.}
		\label{fig_pot_lin}
	\end{center}
\end{figure}

Figure \ref{fig_pot_lin} contains the potentials (a) and cross sections (b) at $\ecm=5.5$ MeV.  The conclusions are similar to those 
of Fig.\ \ref{fig_pot_lip}.  Breakup effects have an important role around $R\sim 4-6$ fm.  The KD global potential provides 
a similar cross section, but only predicts one minimum.  The equivalent potential gives a cross section very close
to the original CDCC calculation.

\section{The $\lid$ scattering}
The dipole resonance observed in the $\lip$ inelastic cross sections \cite{TKA17} was first suggested in a $\lid$ experiment \cite{KST15}.  
In addition to the broad nature of this dipole resonance, which makes theoretical models rather complicated, the low breakup 
threshold of the deuteron requires a $3+2$ model.  The principle of the CDCC formalism remains unchanged with respect to $\lip$ 
($3+1$ model) but the calculations are much longer since $(i)$ the coupling potentials involve multidimension integrals
(see Appendix), $(ii)$ the 
number of channels in the coupled system (\ref{eq20}) is the product of the numbers of pseudostates in $\oli$ and in $d$.  Having a full convergence of the cross sections, in a wide angular range, is therefore a challenge.

\begin{figure}[htb]
	\begin{center}
		\epsfig{file=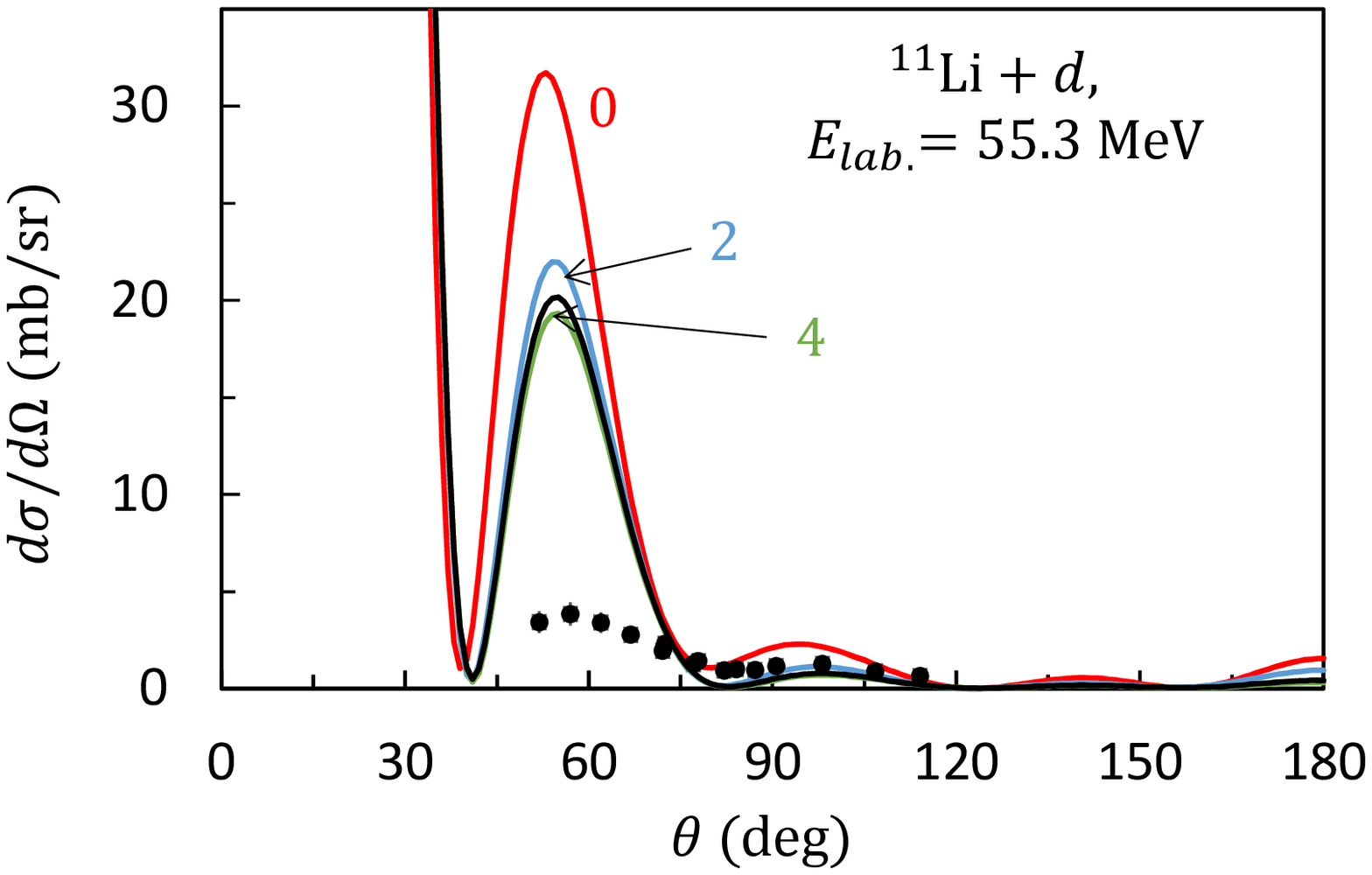,width=7.5cm}
		\caption{$\lid$ elastic cross section at $\elab=55.3$ MeV with the $\lip$ and $\lin$ equivalent
			potentials. The labels indicate the $\jmax$ value in the deuteron. The experimental data are taken from Ref.
			\cite{KST15}.}
		\label{fig_lid1}
	\end{center}
\end{figure}

I have started this exploratory study by using a conventional CDCC approach, where the breakup of $\oli$ is simulated by the 
equivalent $\lip$ and $\lin$ potentials defined in Sec.\ IV.C.  In this way, I deal with a standard CDCC calculation which 
only includes deuteron pseudostates.  The cross section is presented in Fig.\ \ref{fig_lid1}, with the experimental 
data of Ref.\ \cite{KST15}.  As for $\lip$, I adopt a linear scale to highlight the convergence of the calculation. I include deuteron partial waves up to $\jmax=6$, and the maximum energy is $ \emax=20$ MeV.  
At small angles ($\theta \lesssim 30^{\circ}$), the calculation converges rapidly. The data are consistent with a maximum around $\theta \approx 60^{\circ}$, which is supported by theory, but 
the experimental amplitude is lower by a factor 5.  Above $\theta \approx 60^{\circ}$, the experimental oscillation is reproduced 
by the calculation, but the convergence with respect to the deuteron angular momentum is slow.  

\begin{figure}[htb]
	\begin{center}
		\epsfig{file=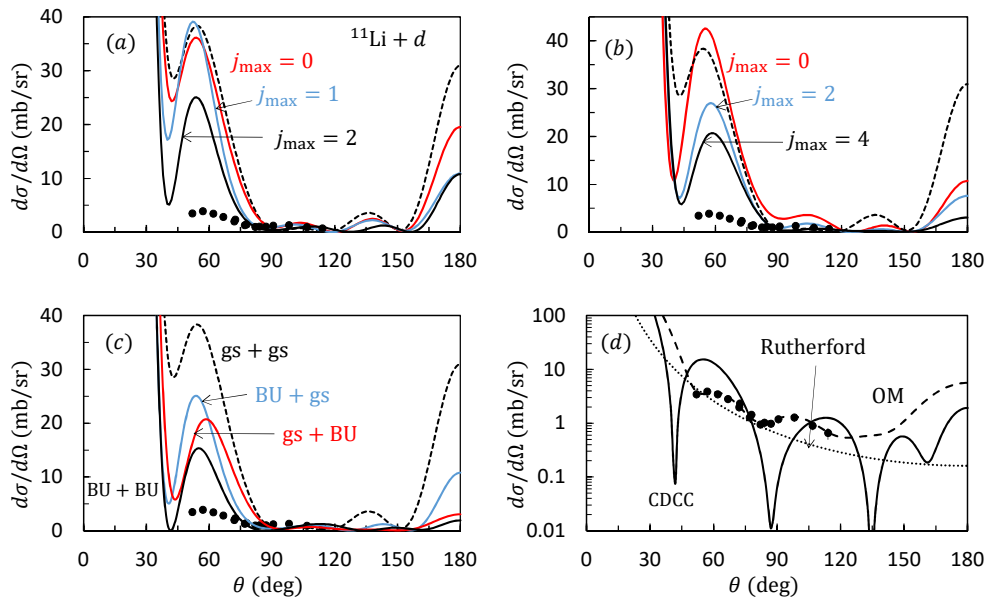,width=13cm}
		\caption{$\lid$ elastic cross sections at $\elab=55.3$ MeV. The experimental data are taken from Ref.\
			\cite{KST15}. (a) Only $\oli$ breakup is included. (b) Only deuteron breakup is included.
		(c) Convergence of the full five-body calculation. (d) Comparison of the CDCC calculation with the Rutherford 
	cross section and with the optical potential of Ref.\ \cite{KST15}. }
		\label{fig_lid2}
	\end{center}
\end{figure}

The full $3+2$ cross sections are displayed in Fig.\ \ref{fig_lid2}.  As these calculations, involving $\oli$ and $d$ breakup simultaneously, 
are extremely time consuming, I first consider single breakup.  Figure \ref{fig_lid2}(a) includes $\oli$ breakup only, 
the deuteron remaining 
in the ground state.  Again, the calculation predicts a maximum near $\theta \approx 60^{\circ}$, but the 
amplitude is overestimated.  As for $\lip$, the role of $j=1^-$ is minor, but $j=2^+$ PS significantly modify
the cross section.

In Fig.\ \ref{fig_lid2}(b), the $\oli$ breakup is neglected, and partial waves up to $\jmax=4$ are included in the deuteron.  
Clearly, increasing $\jmax$ reduces the amplitude of the peak but the calculation still overestimates the data.  
Figure \ref{fig_lid2}(c) illustrates the various possibilities.  When breakup effects are included in $\oli$ and in $d$, 
the amplitude is reduced, but the small experimental values around $\theta \approx 60^{\circ}$ cannot be reproduced.  
Of course, the convergence is not fully achieved.  To keep calculations within reasonable limits, I 
have set $\jmax=2$ for $\oli$, 
and $\jmax=4$ for the deuteron.  With these conditions, the number of $\lid$ states is 1100, and the size of the coupled-channel system (\ref{eq20}) is close to 9000 when the channel spin $I$ and the orbital momentum $L$ are taken into account.  At the moment, it is 
virtually impossible to go beyond these values,  but increasing the CDCC basis might slightly reduce the amplitude of the peak.

\onecolumngrid

\twocolumngrid

\begin{figure}[htb]
	\begin{center}
		\epsfig{file=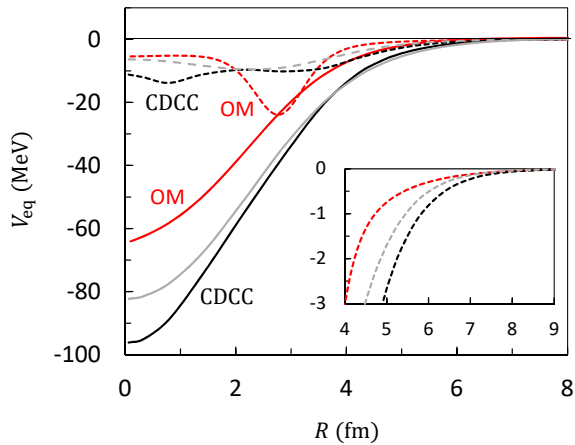,width=7.5cm}
		\caption{Real (solid line) and imaginary (dashed line) $\lid$ equivalent potential compared to the optical potential of Ref.\ \cite{KST15}. The grey lines correspond to the single-channel calculation. The inset presents a zoom on the imaginary
		potential at large distances.}
		\label{fig_lid_pot}
	\end{center}
\end{figure}

In order to analyze the CDCC results, I show in Fig.\ \ref{fig_lid2}(d) (logarithmic scale)
the Rutherford cross section, and the cross section 
computed with the optical potential of Ref.\ \cite{KST15}.  Surprisingly the experimental data are close to 
a pure Rutherford scattering.  The optical potential of Ref.\ \cite{KST15} nicely reproduces the data.  
This potential is compared to the 
CDCC equivalent potential in Fig.\ \ref{fig_lid_pot}.  The main difference between them is that the CDCC predicts a larger range 
for the real and imaginary parts.  In the single-channel approximation (grey lines), the range of the imaginary
part is slightly smaller than in the full calculation. In a reaction involving two fragile nuclei, it seems natural that their interaction extends 
to large distances.  However the optical potential which fits the data is characterized by a fairly short range.  
This apparent contradiction certainly deserves more experimental studies, in particular at other scattering energies.

\section{Conclusion}

The main goal of this work is the simultaneous investigation of $\lip$ and $\lid$ scattering at low energies. 
$\oli$ and the deuteron have a low breakup threshold, which makes the continuum quite important.  In both reactions, 
I have used the same three-body model for $\oli$.  I paid a special attention to E1 transitions and I extended 
a previous calculation of the E1 distribution \cite{PDB12} by considering a correction to the long-wavelength 
approximation.  This correction could give rise to isoscalar E1 transitions, as suggested in Refs.\ \cite{TKA17,KST15} 
on the basis of a large $\oli$ radius.  However, if I confirm the existence of a dipole resonance at low energies, 
the isoscalar part of the E1 matrix element is negligible, owing to the low photon energies.

For the $\lip$ elastic scattering, the CDCC calculation reproduces the experiment fairly well.  The role of 
the $\linn$ continuum can been seen at large angles $(\theta > 90^{\circ})$, small angles being weakly sensitive.  
Notice that the calculations depend on a $\lipb$ potential, which is available from global parametrizations only.  
Experimental data on $\lipb$ elastic scattering would be helpful to derive a more accurate optical potential.

From the CDCC formalism, I have determined $\lip$ and $\lin$ equivalent potentials.  The goal is to use 
them in $\lid$ scattering with the additional deuteron breakup.  The comparison of these equivalent potentials 
with the global potential of Ref.\ \cite{KD03} shows that the main difference is in the range.  This result is 
not surprising since the large radius of $\oli$, as well as its low binding energy, are not considered in 
global parametrizations.

The theoretical description of $\lid$ is a difficult challenge.  The incident energy of $\oli$ is almost the same as 
in $\lip$ (55.3 MeV) and the c.m. energy is therefore almost double.  In spite of this, the $\lid$ data are compatible 
with a pure Rutherford scattering, as shown in Fig.\ \ref{fig_lid2}(d).  I have investigated the $\lid$ system in two ways: 
in the former, I use a standard three-body model using $\lip$ and $\lin$ interactions, and in the latter I extend 
the CDCC formalism to five bodies, with $\oli$ described as $\linn$.  Both approaches provide qualitatively similar 
cross sections.  At small angles $(\theta \lesssim 60^{\circ})$, the apparent peak in the data is present in the 
calculation, but its amplitude is much larger.  Increasing the number of pseudostates reduces the amplitude, 
but it remains overestimated.  

The five-body calculation is a numerical challenge, owing to the very large 
number of channels.  It is difficult to get a perfect convergence although my calculation should not be far 
from convergence.  Of course such calculations have shortcomings: (i) the $\lipb$ and $\linb$ optical potentials are not experimentally known, (ii) antisymmetrization effects between the neutrons of $\oli$ and of the deuteron are neglected, 
(iii) at these low energies, the $\lipb$ and $\linb$ Pauli forbidden states may play a role.

On the experimental side, data so close to Rutherford scattering are unexpected.  The authors of Ref.\ \cite{KST15} fit 
these data with an optical potential presenting a short range.  This is illustrated in Fig.\ \ref{fig_lid_pot}, where 
I compare the CDCC equivalent potential with the optical model of Ref.\ \cite{KST15}.  As for $\lipb$ scattering, 
more data on $\lid$, especially at small angles, would be welcome to confirm the short-range of the $\lid$ potentials.

\section*{Acknowledgments}
I am grateful to R. Kanungo for useful discussions about the experimental data.
This work was supported by the Fonds de la Recherche Scientifique - FNRS under Grant Numbers 4.45.10.08 and J.0049.19.
It benefited from computational resources made available on the Tier-1 supercomputer of the 
F\'ed\'eration Wallonie-Bruxelles, infrastructure funded by the Walloon Region under the grant agreement No. 1117545. 

\onecolumngrid
\appendix
\section{Calculation of the coupling potentials}
\label{appendixa}
The calculation of the coupling potentials (\ref{eq21}) is briefly presented in this Appendix.  I consider systems 
made of a three-body projectile and, either of a structureless target (referred to as $"3+1"$), or of a two-body 
target (referred to as $"3+2"$).  Calculations associated with 2+2 systems have been described in Ref.\ \cite{De18}.

The matrix elements (\ref{eq21}) involve 3 or 6 terms for 3+1 and 3+2 systems, respectively.  I consider one of this 
potentials, $V_{11}$, associated with the Jacobi coordinates of Fig.\ \ref{fig_conf}.  Consequently I 
consider $V_{11}(\pmb{R}+\alpha \pmb{y}_1)$ for 3+1 systems and $V_{11}(\pmb{R}+\alpha \pmb{y}_1+\beta \pmb{r}_2)$ 
for 3+2 systems.  Coefficients $\alpha$ and $\beta$ are related to the masses of the fragments.  This potential is 
expanded in multipoles: for 3+1 systems, I have 
\begin{eqnarray}
V_{11}(\pmb{R}+\alpha \pmb{y}_1)=\frac{1}{\sqrt{4\pi}}
\sum_{\lambda} 
V_{\lambda}(R,y_1) Y_{\lambda}^0(\Omega_{y_1}),
\label{eqa1}
\end{eqnarray}
whereas 3+2 potentials are expanded as
\begin{eqnarray}
V_{11}(\pmb{R}+\alpha \pmb{y}_1+\beta \pmb{r_2})=\frac{1}{\sqrt{4\pi}}
\sum_{\lambda \lambda_1 \lambda_2} 
V_{\lambda \lambda_1 \lambda_2}(R,r_1,r_2)
\bigl[Y_{\lambda_1}(\Omega_{y_1})\otimes Y_{\lambda_2}(\Omega_{r_2}) \bigr]^{\lambda}_0.
\label{eqa2}
\end{eqnarray}
In these expansions, I assume that the $z$ axis is along the $\pmb{R}$ direction.
The multipole components $V_{\lambda}$ and $V_{\lambda \lambda_1 \lambda_2}$ can be computed by numerical 
integration of the potentials over the angles.

Then, inserting expansion (\ref{eqa1}) in the coupling potentials (\ref{eq21}) provide, for 3+1 systems
\begin{eqnarray}
V^{J\pi}_{cc'}(R)=\sum_{\lambda}C^{J\pi (\lambda)}_{Lj_1,L'j'_1}
\sum_{\gamma \gamma'}D^{j_1 j'_1 (\lambda)}_{\gamma \gamma'}
\sum_{K K'}F^{j_1 j'_1 (\lambda)}_{\gamma K, \gamma' K'}(R),
\label{eqa3}
\end{eqnarray}
where coefficients $C^{J\pi (\lambda)}_{LI,L'I'}$ are given by
\begin{eqnarray}
C^{J\pi (\lambda)}_{LI,L'I'}=(-1)^{I'+L+J}\hat{I}\hat{L} \hat{\lambda}^{-1}
\begin{Bmatrix}
I & L & J  \\ 
L' &  I' & \lambda
\end{Bmatrix}
\langle Y_L \Vert Y_{\lambda} \Vert Y_{L'} \rangle,
\label{eqa4}
\end{eqnarray}
and are common to all CDCC calculations, independently of the projectile and target descriptions (in the case
of a structureless target the channel spin is $I=j_1$). I use the notation $\hat{x}=\sqrt{2x+1}$.

Coefficients $D^{j_1 j'_1 (\lambda)}_{\gamma \gamma'}$ are typical of 3+1 systems and are defined as
\begin{eqnarray}
D^{j_1 j'_1 (\lambda)}_{\gamma \gamma'}=
(-1)^{\ell+S+j'_1+\lambda}\, \hat{\ell}\hat{j}'_1
\begin{Bmatrix}
j_1 & \ell & S  \\ 
\ell' &  j'_1 & \lambda
\end{Bmatrix}
\langle 
\bigl[Y_{\ell_x}(\Omega_{x_1})\otimes Y_{\ell_y}(\Omega_{y_1}) \bigr]^{\ell} \Vert Y_{\lambda}(\Omega_{y_1}) \Vert
\bigl[Y_{\ell'_x}(\Omega_{x_1})\otimes Y_{\ell'_y}(\Omega_{y_1}) \bigr]^{\ell'}
\rangle
\delta_{\ell_x \ell'_x}\delta_{SS'},
\label{eqa5}
\end{eqnarray}
where the condition $\ell_x=\ell'_x$ arises from the choice of the Jacobi coordinates.  
Functions $F$ are also associated with 3+1 systems; they are given by
\begin{eqnarray}
F^{j_1 j'_1 (\lambda)}_{\gamma K, \gamma' K'}(R)=
\iint \chi^{j_1}_{\gamma K}(\rho) \chi^{j'_1}_{\gamma' K'}(\rho)
V_{\lambda}(R,\rho \sin \alpha) \Phi^K_{\ell_x \ell_y}(\alpha) \Phi^{K'}_{\ell'_x \ell'_y}(\alpha)
\cos^2 \alpha \sin^2 \alpha \, d\alpha \, d\rho,
\label{eqa6}
\end{eqnarray}
where $\Phi^K_{\ell_x \ell_y}(\alpha)$ are functions depending on the hyperangle 
(see Refs.\ \cite{DB10,De18} for more information).
In this expression, the quadrature over the hyperangle $\alpha$ is performed numerically.  The hyperradial 
functions $\chi^{j_1}_{\gamma K}(\rho)$ are expanded over Lagrange functions, and the integration over $\rho$ is 
therefore straightforward.

Equation (\ref{eqa3}) is associated to a specific term ($V_{11}$) of the total potential $V_{11}+V_{21}+V_{31}$.  I compute the 
matrix elements of $V_{21}$ and $V_{31}$ by using other choices of the Jacobi coordinates.  The corresponding coupling potentials 
are therefore computed with (\ref{eqa3}) followed by a transformation using the Raynal-Revai coefficients \cite{RR70}.  
This method was already adopted in Refs.\ \cite{RAG08,DDC15}.

The extension of 3+2 systems is new.  It can be extended from the previous equation in a systematic way.  In that case, 
the target has a two-body structure and its wave function is given by (\ref{eq16}).  The coupling potentials (\ref{eqa3}) are generalized as
\begin{eqnarray}
V^{J\pi}_{cc'}(R)=\sum_{\lambda}C^{J\pi (\lambda)}_{LI,L'I'}
\sum_{\gamma \gamma' \lambda_1 \lambda_2}\bar{D}^{j_1 j_2, j'_1 j'_2 (\lambda  \lambda_1 \lambda_2)}_{\gamma \ell_2, \gamma' \ell'_2}
\sum_{K K'}\bar{F}^{j_1 j_2, j'_1 j'_2 (\lambda  \lambda_1 \lambda_2)}_{\gamma K \ell_2, \gamma' K'\ell'_2}(R),
\label{eqa7}
\end{eqnarray}
where coefficients $\bar{D}$ are given by
\begin{eqnarray}
\bar{D}^{j_1 j_2, j'_1 j'_2 (\lambda  \lambda_1 \lambda_2)}_{\gamma \ell_2, \gamma' \ell'_2}=
\hat{j_1}\hat{j_2}\hat{I'}\hat{\lambda}
\begin{Bmatrix}
j_1 &j_2 & I  \\ 
\lambda_1 & \lambda_2 & \lambda \\
j'_1 &  j'_2 & I'
\end{Bmatrix}
\langle 
\bigl[Y_{\ell_2}\otimes \chi_s \bigr]^{j_2} \Vert Y_{\lambda_2} \Vert
\bigl[Y_{\ell'_2}\otimes \chi_s \bigr]^{j'_2}
\rangle
D^{j_1 j'_1 (\lambda_1)}_{\gamma \gamma'}.
\label{eqa8}
\end{eqnarray}
The functions $\bar{F}$ are now defined as
\begin{eqnarray}
&& \bar{F}^{j_1 j_2, j'_1 j'_2 (\lambda  \lambda_1 \lambda_2)}_{\gamma K \ell_2, \gamma' K'\ell'_2}(R)=
\iiint \chi^{j_1}_{\gamma K}(\rho) g^{j_2}_{\ell_2}(r_2)
V_{\lambda \lambda_1 \lambda_2}(R,\rho \sin \alpha,r_2) \nonumber \\ 
&&\hspace{1cm}\times \chi^{j'_1}_{\gamma' K'}(\rho)g^{j'_2}_{\ell'_2}(r_2)
\Phi^K_{\ell_x \ell_y}(\alpha) \Phi^{K'}_{\ell'_x \ell'_y}(\alpha)
\cos^2 \alpha \sin^2 \alpha\, d\alpha \, d\rho\, dr_2 ,
\label{eqa9}
\end{eqnarray}
which means that an additional integration over $r_2$ is required.  Again, the two-body radial functions $ g^{j_2}_{\ell_2}(r_2)$ are 
expanded over a Lagrange mesh, and the associated quadrature is simple.  As for 3+1 systems, the use of Raynal-Revai 
coefficients permits a similar calculation for the other components of the potential $V_{21},V_{31},V_{22}$, and
$V_{32}$.  An extension 
to 3+3 systems is feasible by adopting the same technique, and by using a double Raynal-Revai transformation.

\twocolumngrid
%

\end{document}